\documentclass[twocolumn,trackchanges]{aastex701}

\usepackage{xspace}
\usepackage{amsmath}
\usepackage{comment}
\usepackage{rotating}

\newcommand{\ha}{H$\alpha$}
\newcommand{\hb}{H$\beta$}

\newcommand{\lya}{Ly$\alpha$}

\newcommand{\G}{{\rm G}}
\newcommand{\K}{{\rm K}}

\newcommand{\nii}{[N\thinspace{\sc ii}]}

\newcommand{\oiii}{[O\thinspace{\sc iii}]}
\newcommand{\oiiib}{O\thinspace{\sc iii}]}
\newcommand{\oii}{[O\thinspace{\sc ii}]}

\newcommand{\sii}{[S\thinspace{\sc ii}]}

\newcommand{\heii}{He\thinspace{\sc ii}}

\newcommand{\ciii}{C\thinspace{\sc iii}]}

\newcommand{\civ}{C\thinspace{\sc iv}}

\newcommand{\nv}{N\thinspace{\sc v}}

\begin{document}

\title{MOCSS: Multi-Object Coronagraphy for JWST/NIRSpec Slitless Spectroscopy}

\author[orcid=0009-0002-9932-4461]{Mason S. Huberty}
\affiliation{Minnesota Institute for Astrophysics, University of Minnesota, 116 Church Street SE, Minneapolis, MN 55455, USA}
\email[show]{huber458@umn.edu}  

\author[0000-0002-9136-8876]{Claudia Scarlata}
\affiliation{Minnesota Institute for Astrophysics, University of Minnesota, 116 Church Street SE, Minneapolis, MN 55455, USA}
\email{mscarlat@umn.edu} 

\author[0000-0001-8587-218X]{Matthew J. Hayes}
\affiliation{Stockholm University, Department of Astronomy and Oskar Klein Centre for Cosmoparticle Physics, AlbaNova University Centre, SE-10691, Stockholm, Sweden}
\email{matthew@astro.su.se} 

\author[0000-0002-6586-4446]{Alaina Henry}
\affiliation{Space Telescope Science Institute, 3700 San Martin Drive, Baltimore, MD 21218, USA}
\email{ahenry@stsci.edu} 
\author[0000-0002-4153-053X]{Danielle A. Berg}
\affiliation{Department of Astronomy, University of Texas at Austin, 2515 Speedway, Austin, TX 78712, USA}
\email{daberg@austin.utexas.edu}

\author[0000-0002-2070-9047]{Vieri Cammelli}
\affiliation{Department of Physics, Informatics \& Mathematics, University of Modena \& Reggio Emilia, via G. Campi 213/A, 41125, Modena, Italy}
\email{vieri.cammelli@unimore.it}

\author[0000-0001-7782-7071]{Richard Ellis}
\affiliation{Department of Physics and Astronomy, University College London, Gower Street, London WC1E 6BT, UK}
\email{richard.ellis@ucl.ac.uk}

\author[0000-0002-6790-5125]{Anne E. Jaskot}
\affiliation{Department of Physics and Astronomy, Williams College, Williamstown, MA 01267, USA}
\email{08aej@williams.edu} 

\author[0000-0003-2589-762X]{Matilde Mingozzi}
\affiliation{AURA for ESA, Space Telescope Science Institute, 3700 San Martin Drive, Baltimore, MD 21218, USA}
\email{mmingozzi@stsci.edu} 

\author[0000-0003-2083-7564]{Pierluigi Monaco}
\affiliation{Dipartimento di Fisica - Sezione di Astronomia, Università di Trieste, Via Tiepolo 11, 34131 Trieste, Italy}
\affiliation{INAF-Osservatorio Astronomico di Trieste, Via G. B. Tiepolo 11,
34143 Trieste, Italy}
\email{pierluigi.monaco@inaf.it}

\author[0000-0001-8419-3062]{Alberto Saldana-Lopez}
\affiliation{Stockholm University, Department of Astronomy and Oskar Klein Centre for Cosmoparticle Physics, AlbaNova University Centre, SE-10691, Stockholm, Sweden}
\email{alberto.saldana-lopez@astro.su.se} 

\author[orcid=0000-0002-3389-9142]{Jonathan C. Tan}
\affiliation{Dept. of Space, Earth \& Environment, Chalmers University of Technology, Chalmersgatan 4, 412 96 Gothenburg, Sweden} 
\affiliation{Department of Astronomy, University of Virginia, 530 McCormick Rd, Charlottesville, VA 22904, USA}
\email{jctan.astro@gmail.com} 

\author[0000-0001-9136-3701]{Alice Young}
\affiliation{Stockholm University, Department of Astronomy and Oskar Klein Centre for Cosmoparticle Physics, AlbaNova University Centre, SE-10691, Stockholm, Sweden}
\email{alice.young489@gmail.com}

\begin{abstract}

We present an experimental JWST/NIRSpec observational method to obtain $\rm R\gtrsim1000$ G140M/F100LP slitless spectroscopic observations. This ``Multi-Object Coronagraphy for Slitless Spectroscopy" (MOCSS) technique involves opening essentially all microshutters on NIRSpec simultaneously while closing shutters corresponding on-sky with known bright objects. Through observations of the Hubble Ultra Deep Field, we demonstrate the discovery potential of these deep, high resolution observations without the need for target pre-selection. 
We obtain a catalog of 22 secure multi-line emitters at $1.3<z<4.6$, with observations of \oii, \oiii, \ha, \hb\ and \nii\ lines, accounting for $\sim80\%$ of the emission lines identified directly in the dispesed images. The MOCSS survey provides first time spectroscopic redshifts for 18 of these sources. We also demonstrate that first order traces of the NIRSpec/G140M/F100LP observing mode extend substantially beyond the nominal $\sim1.89\rm \mu m$ red cutoff, as far as $2.94\rm \mu m$ in the most extreme case. Compared to existing NIRISS WFSS of similar exposure times, we find that the MOCSS technique can observe very faint lines while covering the full $1-2 \rm \mu m$ regime without gaps in wavelength coverage.
This study describes the highest-resolution slitless spectroscopy campaign of the $1-2 \rm \mu m$ regime and emphasizes the discovery potential of this unorthodox observational technique. 

\end{abstract}

\section{Introduction}\label{section1}
Near infrared (NIR) spectroscopy has been one of the defining successes of the James Webb Space Telescope (JWST). The ability to capture optical emission lines around cosmic noon ($1\lesssim z \lesssim 4$) in the $1-2 \rm \mu m$ wavelength regime offers insights into the conditions of the interstellar medium, galaxy evolution, star-formation, metallicity, the baryon cycle, dust, and emission line luminosity functions \citep[e.g.,][]{henry2021, sanders2021,shapley2023,boyett2024,clarke2026,pang2026}. At higher z ($\gtrsim6$), ultraviolet (UV) emission lines are instead redshifted to this wavelength range, and can be used to explore galaxy, active galactic nuclei (AGN), and intergalactic medium (IGM) conditions in the early Universe \citep[e.g.,][]{vidalgarcia2017,byler2018,kewley2019,llerena2022,mingozzi2024,hayes2025}. 
$\rm R\gtrsim1000$ spectroscopy provides the ability to resolve the fine spectral details that carry this cosmic information, such as closely spaced emission lines, even if at the expense of the continuum sensitivity. It also allows for the detection of emission lines with very low equivalent widths \citep[EWs, e.g.,][]{glazer2025,umeda2026}. 

Wide-field slitless spectroscopy (WFSS) of the NIR is a powerful observing mode of JWST, providing spectra for thousands of objects in the field of view simultaneously, and allowing for the discovery of new sources \citep{treu2022,matthee2023,oesch2023,malkan2025,sun2025}. WFSS also allows for blind emission line searches, without relying on targeting assumptions \citep{martin2008,rauch2008,dressler2011,henry2012,kashino2023,runnholm2025}. The downside of slitless spectroscopy is the potential for overlapping spectral traces, but this can be mitigated when observing at multiple orientations \citep[e.g.,][]{pirzkal2018,watson2025,huberty2026}.
Unfortunately, high-resolution slitless spectroscopy in the $1-2\rm \mu m$ regime remains elusive: JWST NIRCam WFSS provides $\rm R\sim1100-1600$ spectra at $>2.4\rm \mu m$, at too high of a wavelength range to capture most of the optical lines at $z<2.5$ and UV lines at $z\gtrsim7$. JWST NIRISS provides $1-2\rm \mu m$ WFSS coverage, but only at $\rm R\gtrsim150$. The Euclid space telescope's NISP instrument \citep{nisp2025} and the Roman space telescope's WFI \citep{akeson2019} observe the NIR at $\rm R=\sim 500$. 

JWST NIRSpec's medium and high spectral resolution configurations do observe at $\rm R\gtrsim 1000$. However, NIRSpec's multi-object spectroscopy (MOS) observing mode involves isolating pre-selected targets with its micro-shutter assembly (MSA), meaning NIRSpec/MOS observations will not find new sources. As a result, the process of making MSA configurations often leads to survey selection functions that are difficult to quantify.
Additionally, the limited size of the MSA shutters may cut off extended regions/emission of a targeted source \citep[e.g.,][]{wisotzki2016,barisic2025,huberty2025}. 
Therefore, we began exploring new observational modes capable of providing deep, high resolution spectra of untargeted sources in the $1-2\rm \mu m$ wavelength range.

In this paper, we present an experimental new JWST/NIRSpec observing mode called ``Multi-Object Coronagraphy for Slitless Spectroscopy" (MOCSS). In this survey, we observed the Hubble Ultra Deep Field \citep[][]{beckwith2006} by opening the vast majority of all shutters on the NIRSpec MSA, to create slitless observations, while closing shutters on top of known bright objects, to mitigate the typical issue of overlapping traces that often comes with WFSS.

This paper is structured as follows: in section~\ref{sec:observations}, we outline the MOCSS observations and new data processing procedure. In section~\ref{sec:identification}, we identify emission lines detected in the MOCSS observations. Section~\ref{sec:sourceidentification} outlines the methods of identifying the sources that host these emission lines. 
Section~\ref{sec:catalog} presents the MOCSS catalog. We discuss the capability of the MOCSS observations in section~\ref{sec:discussion}.
Our conclusions are outlined in section~\ref{sec:conclusion}.

We assume a flat $\Lambda$CDM Planck cosmology \citep{planck2020} and use the AB magnitude system \citep{oke1983}. Forbidden lines are indicated as follows, if presented without wavelength values:
\oii\  $\lambda\lambda3727.09,3729.88$\AA\AA=\oii, 
\oiii\ $\lambda5008.24$\AA=\oiii, \nii\ $\lambda6585.27$\AA=\nii.

\section{Observations and Data}\label{sec:observations}
The Hubble Ultra Deep Field \citep[HUDF,][]{beckwith2006,koekemoer2013,walter2016}, contained within the Great Observatories Origins Deep Survey-South \citep[GOODS-S,][]{giavalisco2004} is known for its exceptional wavelength coverage, with 11-band Hubble Space Telescope (HST) imaging \citep[][]{beckwith2006,bouwens2010,teplitz2013,ellis2013,hayes2024}, JWST NIRCam imaging and spectroscopy (JEMS; \citealp{williams2023}, JADES; \citealp{eisenstein2023}, and FRESCO; \citealp{oesch2023}) and NIRISS spectroscopy and imaging \citep[NGDEEP,][]{bagley2024}. 

\begin{figure*}
    \centering
    \includegraphics[width=1\linewidth]{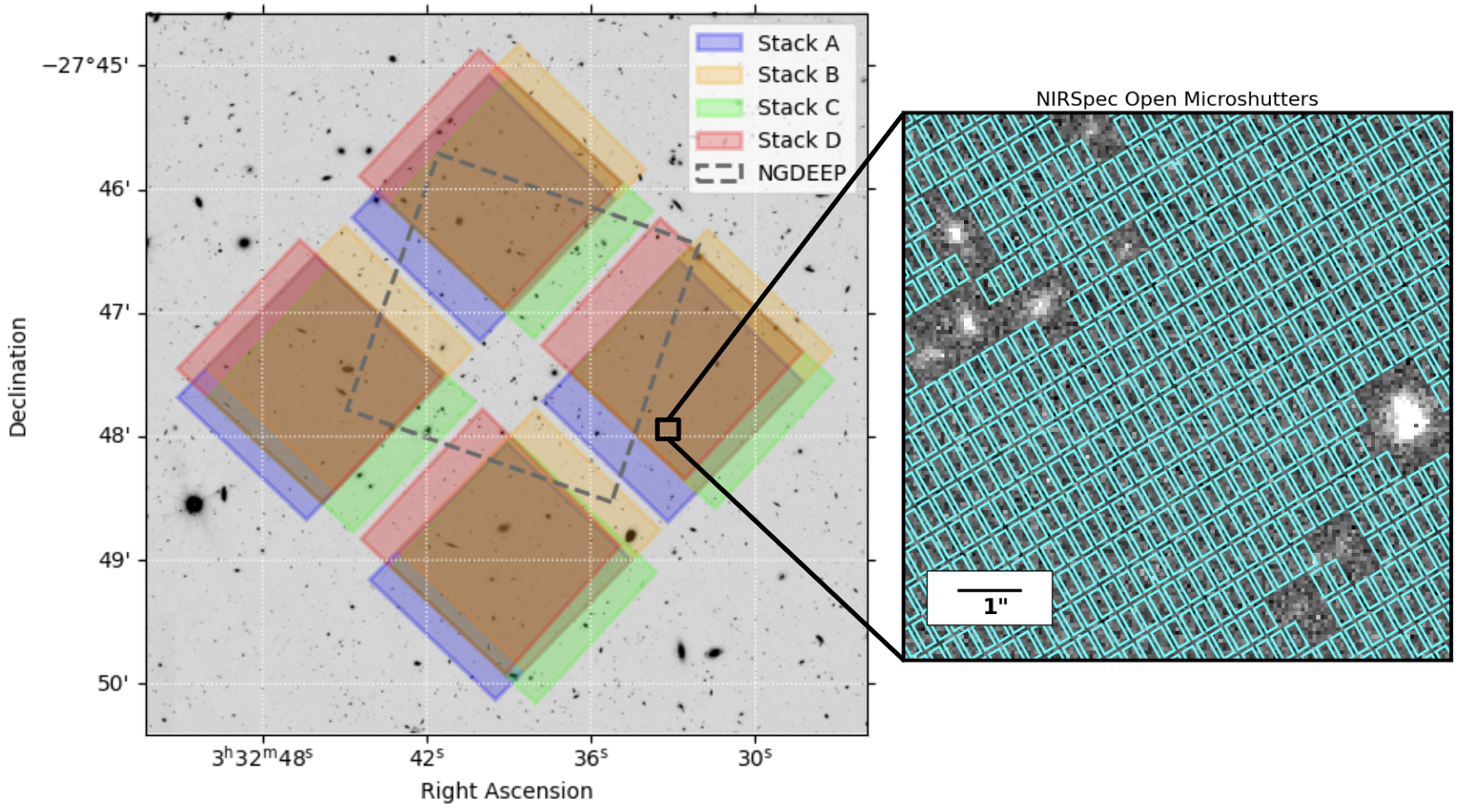}
    \caption{The footprint of the MOCSS survey, divided into the four stacks. The field of view of NGDEEP NIRISS is denoted by the dashed gray line \citep{bagley2024}. The inset shows an example zoom in of the field where each blue rectangle is an open NIRSpec microshutter in Stack A. The locations without shutter coverage correspond to known sources detected in F814W imaging, showcasing the masking process. The background of the shutters insert is a NIRCam F115W image \citep{decoursey2025}. 
    }
    \label{fig:ditherpattern}
\end{figure*}

The Multi-Object Coronagraphy for Slitless Spectroscopy (MOCSS) survey of the HUDF, JWST-GO-3290 (PIs: Hayes \& Scarlata, doi:\dataset[10.17909/tct8-jr92]{http://dx.doi.org/10.17909/tct8-jr92}), was conducted in JWST Cycle 2. 
MOCSS observations were obtained with the NIRSpec G140M/F100LP disperser/filter pairing, which nominally covers the wavelength range between $\sim0.97$ and $\sim1.89$ $\rm \mu m$ at $\rm R\sim1000$ \citep[][]{ferruit2022}, however, multiple examples of extended first order traces extending beyond $\sim1.89$ $\rm \mu m$ as far as $2.77$ $\rm \mu m$ will be presented in this work. 
Instead of using the standard Multi-Object Spectroscopy (MOS) observing mode, all shutters were opened to enable an effectively slitless spectroscopic campaign. 
To prevent bright objects from dominating the dispersed image and because the goal of the MOCSS survey is to identify faint emission line sources, we close shutters associated with flux detected at $\lambda<9600$\AA, identified from HUDF HST F814W imaging \citep{whitaker2019}. See Appendix~\ref{appenidx:maskpreparation} for a more technical discussion of the shutter masking/coronagraphy used in this program. Figure~\ref{fig:ditherpattern} illustrates the method of opening nearly all shutters, except for those that lie on top of these masked sources. 
Accounting for failed shutters and intentionally closed shutters, $\sim 91 \%$ of the MSA area was open. In a technical sense, perhaps the closest survey to MOCSS is the JADES Dark Horse survey \citep{deugino2026}, which permitted a controlled amount of spectral overlaps in NIRSpec G235M/G395M observations, but still only opened a small fraction of the NIRSpec microshutters. Dark Horse targeted opening 854 shutters, enormous in terms of typical NIRSpec observations, but very small relative to MOCSS observations (854 is $<1\%$ of all NIRSpec shutters).

As is conventional for WFSS, observations were taken at two orientations (offset by $89^\circ$), with 8 dithers per orientation. Each of the 16 observations had an exposure time of $2.33$ hours, for a total exposure time of $37.35$ hours for the whole survey.
The two sets of observations were taken approximately 3 months apart, on 23 August 2023 
(orientation 1) and on 19 November 2023 (orientation 2). The $89^\circ$ offset between orientations means that sources that fall in one quadrant in the NIRSpec MSA in orientation 1 fall primarily one quadrant clockwise from the starting quadrant in orientation 2. 
On each orientation, observations were carried out on two main pointings to cover the gap between the detectors and produce a full, unbroken spectral trace. These were offset by 21\arcsec\ in the dispersion direction.  At each of these two locations we execute a 4-point pattern of $\sim3/4$ shutter, which has two functions: the first is to subsample the pixel scale and allow for reconstruction of the PSF, and the second was to average over bar shadow obscuration and optimize a uniform illumination of the field.  We first computed a dithering pattern using steps of 3/4 shutter (with different steps in x- and y- directions) and fine-tuned this to exact dithers that land at integer+1/4 pixel steps from the previous location. A confirmation image of each dither was taken with the NIRSpec F100LP/MIRROR.

The rate files were retrieved from MAST, which were processed through the JWST pipeline stage 1 detector-level corrections \citep{jwstpipeline2023}.
Each of the 16 dispersed images were row-by-row sky subtracted using a custom algorithm; this was required because the strength of the background in each row of the image varies based on the amount of opened microshutters in that row. 
The 16 exposures were divided into 4 groups of 4 images (called groups A, B, C, and D), where each group is separated by orientation and the larger dither across the chip gap (see Table~\ref{tab:dithers} in Appendix~\ref{appendix:ditherinfo} for group and dither information). 

We created deep dispersed images for each group by aligning the individual images within each group and then taking the pixel by pixel weighted-mean of these aligned images. Technical details about what this entails for each extension of the dispersed image files can be found in Appendix~\ref{appendix:countrate}.
This procedure provides 4 deep dispersed image stacks (2 per orientation) that can be used to identify emission lines.

The full field of view from these 4 stacks can be seen in the left panel of Figure~\ref{fig:ditherpattern}. 
The total field of view for NIRSpec is about 3\farcm6 $\times$3\farcm4, but if the gaps between MSA quadrants are excluded, the NIRSpec field of view with all the shutters open is nominally $\sim$3\farcm22 $\times \sim$2\farcm78=$\sim 8.95\ \rm arcmin^2$. The dither pattern prevents areas of reduced throughput caused by the small bars between microshutters. 
Each NIRSpec quadrant typically has $\sim91\%$ of the shutters commanded open, although we know that various shutters fail both closed and open \citep[though are more likely to fail closed than open, see][]{rawle2022}. Therefore, the field of view for one single stack is $\approx8.1\ \rm arcmin^2$. 

The field of view covered by all 4 stacks is smaller: due to the geometric offset between quadrants in the two orientations and the large shift over the chip gap between stacks in the same orientation, $\approx6.4\ \rm arcmin^2$ are covered by all 4 stacks (see Figure~\ref{fig:ditherpattern}).
This geometric offset also means that an additional $\sim1.5\ \rm arcmin^2$ are unique to an individual stack (for a total of $\sim5.9\ \rm arcmin^2$ of additional field of view coverage at lower depth). An additional $\sim1.7\ \rm arcmin^2$ are covered by 3 out of 4 stacks, and another $\sim0.2\ \rm arcmin^2$ are covered by 2 out of 4 stacks.
Therefore, the MOCSS survey has a total field of view of $\approx14.2\ \rm arcmin^2$ with various levels of coverage.

\begin{figure*}
    \centering
    \includegraphics[width=.9\linewidth]{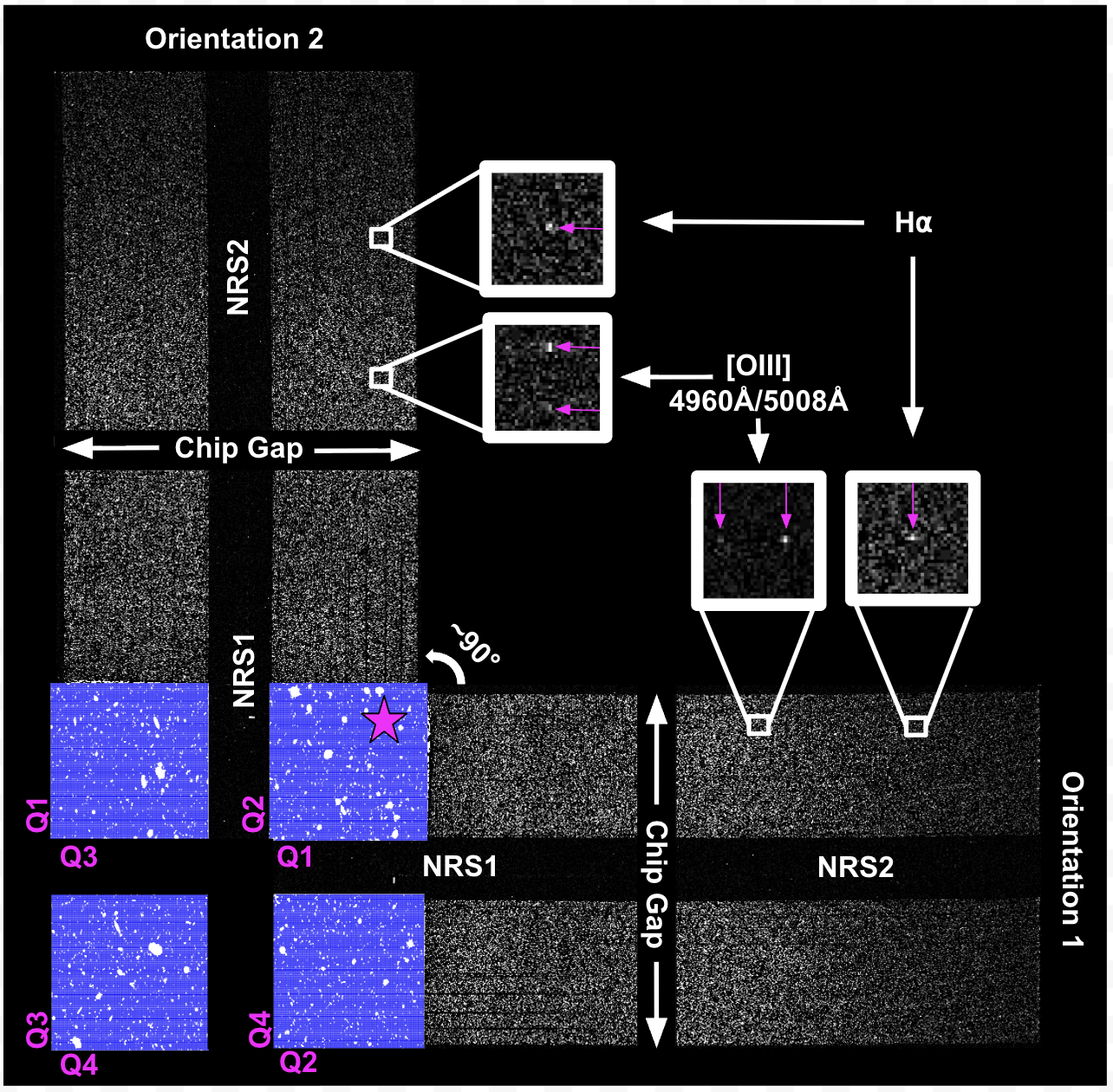}
    \caption{Diagram of the MOCSS observations. The four quadrants of the NIRSpec MSA (labeled Q1 for quadrant 1, etc.) can be seen in the lower left corner, where blue represents open shutters and the white patches signify closed shutters. The corresponding dispersed images for the four quadrants are seen towards the right side of the page for orientation 1, and towards the top of the page for orientation 2. The two detectors are labeled as NRS1 and NRS2, with a chip gap in between. 
    The overlap between the direct images of the NIRSpec quadrants and the detectors demonstrates the overlap between the NRS1 dispersed image with the zero order images of quadrants 1 and 2.
    The location of an example line emitter is marked by the pink star, with the corresponding \ha\ and \oiii\ emission lines identified in both orientations. 
    }
    \label{fig:detectorimage}
\end{figure*}

\section{Emission Line Detection}\label{sec:identification}
In typical slitless spectroscopy, direct images of the field are used to extract the first order spectra. 
In MOCSS, emission lines are instead identified directly in the dispersed images \citep[e.g., in the manner of][]{kashino2023,runnholm2025}.

In each of the four stacks, emission lines were identified using the Python library for Source Extraction and Photometry (\textsc{sep}) package \citep{barbary2016,bertin1996}.
A spatially variable background image was created using \textsc{sep.Background} with $32\times32$ pixel background boxes. 
We retain sources with more than 5 contiguous pixels with flux above $3\times\sigma_{rms}$, the background noise root mean square in the stacked image. Each detector stack has $\sim200-300$ candidate emission lines identified by \textsc{sep} (although it is worth mentioning that many of these sources are artifacts, such as spikes in noise or the edges of the detector).

To remove artifacts we consider the following: within one orientation, real emission lines move between the 2 different stacks (A compared to B or C compared to D) corresponding to the large dither across the chip gap. 
Spatial and wavelength-dependent distortions introduce additional offsets that vary across the field of view. We therefore identify real emission lines in a single orientation if their centroids in both stacks fall within 4 pixels of the expected offset between stacks. This offset requirement may potentially omit real emission lines only covered by 1 stack, but removes a substantial amount of artifacts from the sample.
We also visually inspected the emission line candidates, to ensure that they are not located on the edge of the detector (which can result in edge effects being identified as candidates) or other artifacts. 
This procedure identifies 21 emission lines in orientations 1 and 21 emission lines in orientation 2, for a total of 42 emission lines. 
We will next determine which of these lines belong to the same sources observed in two orientations.

\section{Source Identification}\label{sec:sourceidentification}
Line-emitting sources fall into two categories: those that produce emission lines that are identified in both dispersion orientations, and those that are only identified in one orientation. 

\subsection{Line-Emitters Found in Two Orientations}\label{sec:twoorientsources}
In slitless spectroscopy, without identifying the relevant host, it is not possible to deduce the observed line wavelengths.
This is because the wavelength corresponding to a pixel on the detector image depends on the location of the source in the field of view, and thus a given detector pixel can be associated with different observed wavelengths for different sources. 
This is why the field was observed at two different orientation angles: if an emission line in one orientation can be attributed to the same host and wavelength as an emission line in the other orientation, these lines are likely the same line.

We obtained matching pairs of emission lines between the two orientations with the following procedure: 
(1) For each orientation, we used WCS transformations that are part of the JWST pipeline to map every emission line candidate to every possible MSA shutter that could host the source of that line. Each host shutter corresponds to a different location on the sky and corresponding observed wavelength for the line.
(2) Pairs of lines that map to the same position on the sky (each from a different orientation), within a distance of shutter size (the long side of the shutter, 0\farcs46) and an observed wavelength agreement within 16.6\AA\ (the extracted wavelength difference between two neighboring shutters in the dispersion direction), are considered matches. This method is validated with spectral extraction (see section~\ref{sec:extraction}).
Details on the requirements for running such a orientation matching procedure can be found in Appendix~\ref{appendix:orientationmatchingcode}. 

This procedure is illustrated in Figure~\ref{fig:detectorimage}. The four quadrants of the NIRSpec are displayed in blue, with the corresponding dispersed images on the detectors on the right of the page (orientation 1), and on the top of the page (orientation 2). The direct images of quadrants 1 and 2 of the NIRSpec MSA overlap with the NRS1 detector; this is to visualize the zeroth-order images that appear on the NRS1 detector in MOCSS observations. This results in the substantial loss of about $44\%$ of detector coverage in the NRS1 detector, concentrated on the bluer part of the wavelength range. The highlighted example emission lines (\ha\ and \oiii) can be traced to a single originating point from the coordination of the two dispersed images from the two orientations (in this example, to the location of the pink star). 

We find seven emission lines that are identified by \textsc{sep} in all four stacks/both orientations. One of these pairs of lines is the emission line discussed in \citet{huberty2026mocssline}, which is of interest because it has no corresponding source in deep broadband imaging. The remaining six lines have clear counterparts in deep broadband imaging of the JWST Advanced Deep Extragalactic Survey \citep[JADES][]{eisenstein2023,rieke2023} DR5 \citep{robertson2026,alberts2026}, one of the deepest photometric and spectroscopic campaigns of the HUDF/GOODS-S. These sources are undetected in the F814W, explaining how they were included in the MOCSS configurations, but can be seen at other wavelengths, e.g., in the F150W. 
This orientation matching procedure therefore accounts for 14/42 emission lines identified in section~\ref{sec:identification}, meaning a majority of the lines are not yet accounted for. 

\subsubsection{Line-Emitters Found in 3 Stacks}\label{sec:3stacks}
Emission lines found in 3 of the 4 stacks are likely real, as the observation/dither pattern could miss the feature in one stack.
There are several reasons why a line may not be identified by \textsc{sep} in just one of the four orientations. Firstly, only about $45\%$ of the field of view is covered by all 4 stacks (see Figure~\ref{fig:ditherpattern}). About $12\%$ of the field of view is covered by exactly 3 stacks, meaning if an emission-line source falls in these regions, it would still be identified as a real emission line in one orientation but not the other (a good example of this is MOCSSHUDF3, see Table~\ref{tab:mocsssources}). 
Additionally, the location of a source within the field of view dictates the wavelength coverage of the G140M/F100LP observations. 
For example, many sources in quadrants 3 and 4 and the left side of quadrants 1 and 2 (as seen in Figure~\ref{fig:detectorimage}) have traces that cross the chip gap, where emission lines can fall into the gap in one of the stacks. While the dither pattern is designed to mitigate this effect, it prevents emission lines in this region from initially being identified as a candidate given the expected offset requirement from section~\ref{sec:identification}.
Another reason for differing wavelength coverage between orientations is that for sources in quadrants 3 and 4, the bluest part of the trace can overlap with the zero order image of quadrants 1 and 2 on NRS1, rendering this portion of the trace substantially shallower.

Failed shutters also affect the field of view coverage of the MOCSS strategy: even though the majority of shutters are commanded open, it is known that roughly $7.63\%$ of unvignetted shutters are stuck closed, primarily in quadrants 3 and 4.\footnote{As of October 2024, see \url{https://jwst-docs.stsci.edu/jwst-near-infrared-spectrograph/nirspec-instrumentation/nirspec-micro-shutter-assembly}.}
In typical NIRSpec/MSA observations,  targets can be placed so as to avoid regions of the NIRSpec MSA with failed shutters. In NIRSpec/MOCSS, this is unavoidable. In general, there is also a random failure rate of about $\sim4\%$ for opening a shutter when commanded open \citep{rawle2022}. 
There are also shutters that remain stuck open when ordered to close, although this is a much smaller fraction than those that are stuck closed. 
The zero order contamination and known failed shutter effects reduce the efficacy of MOCSS observations for the field of view that falls within quadrants 3 and 4 of the MSA (relative to sources that fall within the fields of view of quadrants 1 and 2). 

Another factor is that the depth of observations spatially varies across the detector (see section~\ref{sec:depth} for further discussion on the depth of observations). This means faint emission lines may get lost in regions of the detector that do not achieve the same depth as others, and therefore  are not identified by \textsc{sep}.

With all these caveats in mind, we again utilize the orientation matching procedure (see Appendix~\ref{appendix:orientationmatchingcode}) for each remaining emission line. This time however, instead of checking for matches across the two orientations in just the finalized list of 28 remaining emission lines, we look for any matches in the other orientation in the individual \textsc{sep} identified candidates detector images (of which there are 200-300 candidates per stack, prior to being culled by the expected offset criteria from section~\ref{sec:identification}). 
If there is indeed a match to an emission line candidate in this third stack (as two stacks have already identified an emission line in the first orientation), we then visually inspected the predicted location of the line in the fourth stack. If the line falls in the chip gap or off the detector in this fourth stack, this explains why the source was not identified. 
If the line should fall on the detector in the fourth stack, there must be some hint of the line visible through a visual inspection for us to keep the object, even it falls below the level of confidence that \textsc{sep} requires. 
Through this procedure, we identify the hosts for 17 more emission lines, 16 of which have JADES DR5 photometric identification at this point, meaning 31/42 emission lines are now accounted for.
The 17th source was not identified in JADES DR5. However, it is clear in the JADES imaging that there is a source at the expected location (this source was enveloped by the segmentation map of a much larger and brighter neighboring object in the JADES source identification procedure).

\begin{sidewaystable*}[!htpb]
\centering
\caption{The table of sources that have been identified in the MOCSS survey, listed in order of spectroscopic redshift. For all sources identified in the JADES DR5, we list their corresponding JADES ID, $\rm z_{phot}$, and $\rm z_{spec}$, if available. The ``Lines of Interest" column lists both \textsc{sep}- and non-\textsc{sep}- identified lines of interest in each sources spectrum (some lines are too faint/fall off the detector in individual stacks, but are evident in the final spectrum). There may be additional lines of interest not highlighted here. The ``Stacks Identified In" column lists the number of stacks at least one emission line in each source's spectrum was identified in by \textsc{sep} (the maximum value is 4). 
The R.A./Decl. of the sources are taken as the JADES DR5 R.A./Decl. of sources where available. 
The $\dagger$ symbol next to the MOCSS ID number indicates that the lines originate from a source in the `improperly opened' shutter region (see section~\ref{sec:failedopen}). MOCSSHUDF23 is the Ly$\alpha$ emitter described in section~\ref{sec:uvlines}, and the $3^{**}$ in the final column indicates that while the line is seen in 3 stacks, the source was not identified by \textsc{sep}. MOCSSHUDF24$^*$ is the orphan line source from \citet{huberty2026mocssline}, which does not have a secure identity.}\label{tab:mocsssources}
\begin{tabular}{ccccccccc}
\hline
MOCSS ID & R.A. & Decl. & JADES DR5 ID & MOCSS z$\rm _{spec}$ & JADES z$\rm _{phot}$       & JADES z$\rm _{spec}$  & Lines of & Stacks \\ 
& ($\rm ^\circ$) & ($\rm ^\circ$) & & & & & Interest & Identified In \\
\hline
MOCSSHUDF1 & 53.1327744 & -27.7888699 & 119238       & 1.395             & $1.52^{+0.00}_{-0.06}$ & - & \oiii, \ha\ & 2                 \\
MOCSSHUDF2 & 53.1662674 & -27.7559452 & 215316       & 1.550             & $1.54^{+0.00}_{-0.01}$ & -   & \oiii, \ha\ & 4                \\
MOCSSHUDF3 & 53.1350822 & -27.7767391 & 394036      & 1.753             & $1.97^{+0.00}_{-0.20}$ & -  & \oiii, \ha\ & 3                 \\
MOCSSHUDF4 & 53.1596298 & -27.7687187 & 136015       & 1.848             & $1.88^{+0.03}_{-0.09}$ & -      & \oiii, \ha\ & 4             \\
MOCSSHUDF5 & 53.1890869 & -27.7786827 & 208562       & 2.105             & $2.05^{+0.20}_{-0.00}$ & -  & \oiii, \ha\ & 3                 \\
MOCSSHUDF6 & 53.1553192&-27.7637463&140344       & 2.134             & $2.32^{+0.00}_{-0.01}$ & -    & \oiii, \ha & 3               \\
MOCSSHUDF7 & 53.1657257&-27.7563152&145632       & 2.305             & $2.29^{+0.00}_{-0.21}$ & -     & \oiii, \ha\ & 3              \\
MOCSSHUDF8 & 53.1903191&-27.782692&124606       & 2.318             & $2.05^{+0.25}_{-0.01}$ & -     & \oiii, \ha\ & 3              \\
MOCSSHUDF9 & 53.1709061&-27.7754498&209357      & 2.453             & $3.82^{+0.07}_{-0.03}$ & 2.4534    & \oii, \ha, \nii, \sii\    & 4      \\
MOCSSHUDF10 & 53.1862793&-27.7970638&113950       & 2.582             & $2.56^{+0.00}_{-0.02}$ & -     & \oiii, \ha\ & 3              \\
MOCSSHUDF11 & 53.13344&-27.78009&No ID        & 2.645             & -                       & -   & \oii, \oiii\ & 3                \\
MOCSSHUDF12 & 53.1614265&-27.8111496&198545      & 2.824             & $3.25^{+0.08}_{-0.05}$ & 2.8235     & \ha, \nii\ & 3         \\
MOCSSHUDF13 & 53.1605072&-27.7708454&133576       & 2.850             & $2.90^{+0.00}_{-0.02}$ & -     & \oiii & 3              \\
MOCSSHUDF14$\dagger$ & 53.1310921&-27.79039&204595      & 2.869             & $2.90^{+0.82}_{-0.01}$ & -    & \hb, \oiii, \ha\ & 2               \\
MOCSSHUDF15$\dagger$ & 53.1368446&-27.7861919&206126     & 2.874             & $2.92^{+0.12}_{-0.04}$ & 2.8709    & \ha\ & 2          \\
MOCSSHUDF16 & 53.154686&-27.7712822&417234       & 2.947             & $2.95^{+0.49}_{-0.02}$ & -  & \oiii\ & 3                 \\
MOCSSHUDF17 & 53.1603394&-27.7571049&214941       & 3.190             & $3.36^{+0.00}_{-0.16}$ & -   & \oiii\ & 4                \\
MOCSSHUDF18 & 53.1593094&-27.7609806&214036       & 3.322             & $3.20^{+0.13}_{-0.02}$ & -   & \oiii\ & 4                \\
MOCSSHUDF19 & 53.1333046& -27.7803707 &  126242  & 3.474             & $3.59^{+0.09}_{-0.06}$ & 3.4726    &  \hb, \oiii, \ha\ & 3            \\
MOCSSHUDF20 & 53.1605644&-27.7711582&286104       & 3.727             & $3.72^{+0.00}_{-0.03}$ & -  & \oiii &   3               \\
MOCSSHUDF21 & 53.1381683&-27.8012524&111322       & 3.889             & $4.29^{+0.01}_{-0.27}$ & -     & \oiii\  & 3              \\
MOCSSHUDF22$\dagger$ & 53.1619682&-27.8191433&409818      & 4.537             & $4.65^{+0.04}_{-0.06}$ & -  &  \oiii\ &   2            \\
\hline
MOCSSHUDF23 & 53.158905&-27.7650757&213084      & 8.488             & $8.52^{+0.01}_{-0.02}$ & 8.4831 & \lya\ & 3$^{**}$ \\
MOCSSHUDF24$^*$ & 53.13127 & -27.78908 & - & - & - & - & ? & 4 \\
\hline
\end{tabular}
\end{sidewaystable*}

\subsection{Line-Emitters found in 1 Orientation}

Of the 42 identified emission lines, 11 emission lines still have no known origin. 
Without a detection of an emission line in the second orientation, both the host and the identity of an emission line are harder to infer. However, the hosts of multi-line emitters are easier to infer, as the spacing of the lines on the detector permits only certain emission line/redshift combinations. 

The NIRSpec/G140M/F100LP traces are slightly curved in the dispersed images, meaning that if an object is a multi-line emitter, \textsc{sep} will identify multiple emission line candidates at a related y-coordinate in the dispersed images. From the blue end to the red end of the trace in the dispersed images, emission lines belonging to the same host can change by up to $\sim15$ pixels in the cross-dispersion direction.
Therefore, we identify emission lines belonging to the same host as those within $\leq15$ pixels of each other in the y-direction of the dispersed images.
Once the related emission lines are identified, their spacing on the detector can be used to determine the host and identity of the lines.
Additionally, even if an emission line does not appear to be associated with any other emission lines, we can still cut out traces containing this line for all potential hosts and combine the two stacks from the same orientation. This can improve the SNR of weaker lines that were not detected in either stack individually. 
Some of the 31 emission lines that are already accounted for fall into this multi-line category, but we find two additional multi-line sources using this technique. One is an \oiii\ \& \ha\ multi-line emitter initially identified by just the \oiii\ 5008.24\AA\ and \ha\ lines. When the spectra for both stacks in the same orientation are extracted, a weaker but distinct secondary emission spike from the \oiii\ 4960.30\AA\ line appears. This source (MOCSSHUDF1) is only seen in orientation 2, because in orientation 1, the source falls on top of closed rows of microshutters due to electrical shorts in the MSA\footnote{See Figure 5 in \url{https://jwst-docs.stsci.edu/jwst-near-infrared-spectrograph/nirspec-observing-modes/nirspec-multi-object-spectroscopy}}. 

The remaining multi-line source is more complicated: the lines seen in 2 stacks of one orientation are \ha\ and \oiii\ 4960.30\AA\ (the weaker of the \oiii\ doublet). This is a source where the \oiii\ 5008.24\AA\ line falls in the chip gap in one stack, and the \hb\ line falls in the gap in the other stack, emphasizing the importance of the large dither over the chip gap.
This source (MOCSSHUDF14) is peculiar for another reason: according to the MSA metadata files, this source should have been masked by a closed shutter regardless of orientation. This means there is a region where the MSA has multiple shutters that are open even though they were commanded closed (as this source appears in multiple dithers).

\subsubsection{Improperly Opened Shutter Region}\label{sec:failedopen}
Quadrant 4 of NIRSpec in the MOCSS observations has a region where there are several MSA rows of improperly opened/closed shutters that do not match the MSA metadata files for the MOCSS observations.\footnote{Through discussion with the STSCI Help Desk, this appears to be an anomaly with the technical setup of the MOCSS observations.} This behavior is seen in the same region of quadrant 4 in all dithers/orientations of the MOCSS observations. This is confirmed when comparing the MSA metadata files to the confirmation images taken during the actual observations. As such, there is a region in quadrant 4 that allows sources that are bright in the F814W to be observed in the MOCSS survey (i.e., they are not masked as intended). 
Apart from the aforementioned multi-line emitter, an additional 2 emission lines can be traced to sources in this region (MOCSSHUDF15 and MOCSSHUDF22). MOCSSHUDF15 has existing spectroscopy that confirms the legitimacy of these lines. Even though these objects are bright interlopers, their strong lines help confirm the measured redshift of the MOCSS HUDF sources (see section~\ref{sec:redshiftcomparison}).

This means that 35/42 emission lines ($\sim83\%$) are now accounted for, comprising 23 unique emission line sources. 7 lines remain un-characterized, and are discussed in section~\ref{sec:unknown}.

\subsection{Spectral Extraction}\label{sec:extraction}

For lines with an identified host, their spectra can be extracted.
We cannot use the JWST science calibration pipeline \citep{jwstpipeline2023} in its default setting because of the unorthodoxy observing strategy.
However, the pipeline and the associated files can be modified to achieve an appropriate extraction. These modifications are outlined in Appendix~\ref{appendix:pipeline}. 
We use the JWST pipeline \citep{jwstpipeline2023} v1.19 for spectral extraction. We used the jwst\textunderscore 1413.pmap reference file in the JWST Calibration References Data System \citep[CRDS,][]{jwstcrds2016}.

\subsubsection{Extended First Order Emission Lines}\label{sec:extendedfirst}
We must now discuss a very important observation: 16 real emission lines are seen at wavelengths longer than the nominal cutoff at $\rm 1.89\mu m$. 
This means that many emission lines actually originate from extended first order traces. 

The NIRSpec G140M/F100LP disperser/filter pair nominally has an upper wavelength cutoff of about $\sim1.89 \mu \rm m$ \citep{jakobsen2022}. However, while the transmission drops substantially above $\sim1.89 \mu \rm m$, it continues at redder wavelengths, albeit without proper calibrations for spectral extraction. Additionally, this wavelength regime may be contaminated by second order spectra. 

In typical NIRSpec/MSA observations, any extended first order emission lines above $\sim1.89 \mu \rm m$ may go unnoticed, as the pipeline does not extract the uncalibrated, contaminated wavelengths beyond $\sim1.89 \mu \rm m$. However, in the MOCSS observations, it is natural to identify these emission lines in the extended first order trace alongside the emission lines located at $\sim0.97-1.89 \mu \rm m$. 
This is evident from simply looking at the NRS2 detector: some emission lines are identified on the far right side of the NRS2 detector, too far along the detector than would be found for traces limited to nominal G140M/F100LP observations\footnote{See figure 4 in \url{https://jwst-docs.stsci.edu/jwst-near-infrared-spectrograph/nirspec-observing-modes/nirspec-multi-object-spectroscopy}}. 
There is a chance that these lines are second order contaminants, but given that there is no corresponding first order lines for these sources at $\lambda_{\rm observed}/2$ and the redshift comparison analysis for these sources in section~\ref{sec:redshiftcomparison}, this possibility is minute. We also have JADES PRISM and G235M/F170LP spectroscopy for some of these sources that confirms the legitimacy of these emission lines in the extended first order.

Therefore, for any sources with extended first order lines, we must instead use the \textsc{msaexp} manual extraction of NIRSpec MSA spectra \citep{msaexp2022}\footnote{\url{https://github.com/gbrammer/msaexp}}.
\textsc{msaexp} provides spectral extraction as far as $\sim3\rm \mu m$ for G140M/F100LP observations (the reddest line we find is $2.77\rm \mu m$, see section~\ref{sec:catalog}).

\section{The MOCSS Sample}\label{sec:catalog}
We now present the sources identified in the MOCSS survey, which are listed in Table~\ref{tab:mocsssources}.
%lists the sources identified by the emission line analysis of section~\ref{sec:sourceidentification}.
From the 35 accounted for emission lines, 22 line-emitting sources were found \citep[excluding the source from][]{huberty2026mocssline}. Each source was identified in the JADES DR5 photometric catalog, with the exception of one which is blended in the JADES catalog. Four sources also have existing JADES DR4 spectroscopy and spectroscopic redshifts \citep{curtislake2025,scholtz2026}. Table~\ref{tab:mocsssources} also lists the photometric ($\rm z_{phot}$) and spectroscopic ($\rm z_{spec}$) redshifts from JADES, where available.

\begin{figure*}
    \centering
    \includegraphics[width=1\linewidth]{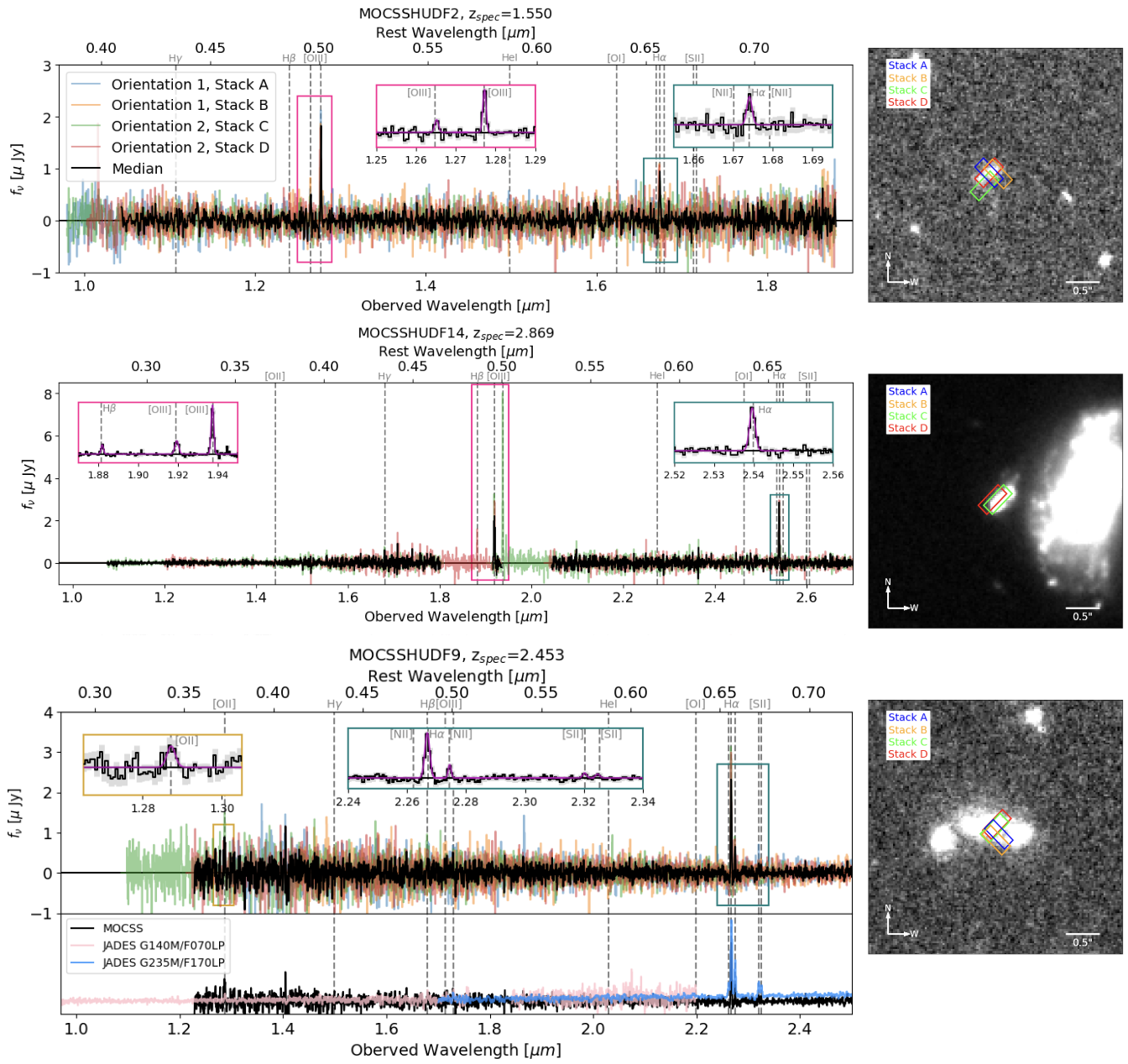}
    \caption{Three example MOCSS spectra for three sources with emission lines seen in both orientations at z=1.550 (top), z=2.869 (middle), and z=2.453 (bottom). 
    The corresponding direct image in JADES F150W images is seen on the right panels, with the corresponding MOCSS MSA shutters used for extraction plotted in blue for stack A, orange for stack B, green for stack C, and red for stack D. 
    The insets show zoom-ins on emission lines of interest: pink insets highlight \oiii\ and \hb\ lines of interest, teal insets highlight \ha\ and \nii\ lines of interest, and gold insets highlight \oii\ lines of interest.
    The source in the top panel is a strong example of an \oiii\ and \ha\ emitter, seen in all four stacks. The source in the middle panel illustrates a source that originates from the failed open shutter region (see section~\ref{sec:failedopen}), meaning the source is only seen in one orientation, but not the other. This source also demonstrates the importance of the dither over the chip gap, as the \hb\ line falls in the chip gap in stack C, while the \oiii\ 5008.24\AA\ line falls in the chip gap in stack D. There is a slight overlap in coverage at the location of the \oiii\ 4960.30\AA\ line in between the two chip gaps. The source in the bottom panel shows a source with existing spectroscopic coverage in JADES (shown in the secondary bottom panel), with strong \ha\ but also \nii, \oii\ and a little \sii\ $\lambda\lambda6718.29,6732.67$\AA\AA\ emission. The complete figure set (23 images) is available in the online journal.
    }
    \label{figset:spectraexamples}
\end{figure*}

Selected spectra are shown in Figure~\ref{figset:spectraexamples}. For each source, the spectra from each of the four stacks are color-coded by stack, and a median of the four stacks is plotted in black. Insets showing prominent emission line features are also included. 
If spectroscopy from JADES is available, we include a secondary panel showing a qualitative comparison between the median MOCSS spectrum and available JADES spectroscopy for that source (see the bottom panel in Figure~\ref{figset:spectraexamples}). We do not necessarily expect the emission lines from JADES and MOCSS to quantitatively match, given that the JADES spectra are aperture-limited by the size of the NIRSpec microshutter, whereas MOCSS is essentially aperture-less given its slitless nature. 
In the right panels of figure~\ref{figset:spectraexamples}, we also show the emission line sources in JADES NIRCam F150W broadband imaging. The nearest reference shutter from each stack (i.e., the relevant shutters from exposures 1, 5, 9, and 13, see Appendix~\ref{appendix:ditherinfo}) to the source is also plotted. Note that in the MOCSS observing mode, for every plotted reference shutter, 3 other smaller dithers about the source also exist, but are not plotted for clarity. In the example source in the middle panel of figure~\ref{figset:spectraexamples}, the galaxy is masked in one stack/orientation (see section~\ref{sec:failedopen}), so only the shutters from stacks where the source is not masked are plotted.

We measure the spectroscopic redshifts from MOCSS from the strong nebular emission lines (always \oiii\ 5008.24 and/or \ha). We find sources between $z=1.395$ and $z=4.537$, with a mean (median) redshift of 2.709 (2.735).

The median spectra per source also emphasize weaker features that were not easily noticeable in the individual stacks, and clear examples of \oiii, \ha, \oii, \nii, \hb, and \sii\ $\lambda\lambda6718.29,6732.67$\AA\AA\ are found. One of the advantages of the MOCSS observation strategy is that at R$\gtrsim1000$, we can easily distinguish the two peaks of \oiii\ and the neighboring \ha\ and \nii\ lines from one another.
 The highest wavelength emission lines that are found is \ha\ at $2.937\rm \mu m$ in MOCSSHUDF19 ($z=3.474$; \ha\ is also spectroscopically confirmed in JADES), followed by \oiii\ at $2.77\rm \mu m$ in MOCSSHUDF22 ($z=4.537$), well past the $1.89\rm \mu m$ nominal upper wavelength cutoff of the G140M/F100LP observations.

\section{Discussion}\label{sec:discussion}

\subsection{Redshift Comparison}\label{sec:redshiftcomparison}
\begin{figure}
    \centering
    \includegraphics[width=.99\linewidth]{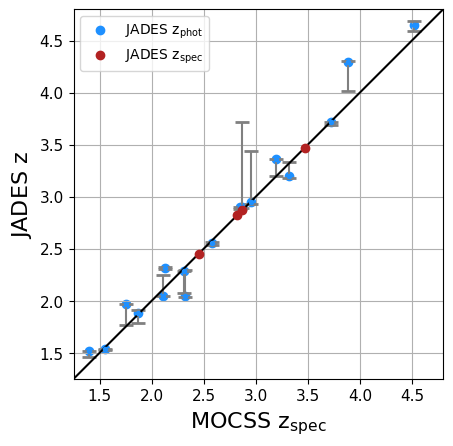}
    \caption{Redshift comparison between the spectroscopic redshifts derived from the MOCSS survey, and the photometric (spectroscopic) redshifts from JADES in blue (red). The one-to-one line is shown in black. The two surveys are in strong redshift agreement. 
    }
    \label{fig:redshiftcomparison}
\end{figure}

We now compare the redshifts measured in MOCSS with those measured in JADES. Figure~\ref{fig:redshiftcomparison} shows this comparison, with the spectroscopic redshifts from MOCSS on the x-axis and the JADES redshifts on the y-axis. If spectroscopic redshifts from JADES exist, the JADES redshift is taken as the spectroscopic redshift (red points). Otherwise, the photometric redshifts are used (blue points). Both the spectroscopic and photometric redshifts from JADES are strongly consistent with those from MOCSS. 
To quantify this, we compute the normalized median absolute deviation, $\sigma_{NMAD}$, between the MOCSS $\rm z_{spec}$ and the JADES $\rm z_{phot}$ (disregarding sources that have $\rm z_{spec}$ measurements from both MOCSS and JADES) as follows:
\begin{equation}
    \sigma_{NMAD}=1.48\times \rm median \bigg( \bigg| \frac{\Delta z}{1+z} - median \bigg(\frac{\Delta z}{1+z} \bigg) \bigg| \bigg)
\end{equation}
where $\frac{\Delta z}{1+z} = \frac{z_{phot}-z_{spec}}{1+z_{spec}}$, in the manner of \citet{hoaglin1983,brammer2008,barro2011,barro2019,casey2023,rieke2023,wang2024,ratajczak2026,huberty2026,robertson2026}. 
We measure a $\sigma_{NMAD}=0.030$, indicating a strong agreement between the JADES photometric and MOCSS spectroscopic measurements of the redshifts and emphasizes the credibility of the redshifts measured in MOCSS.

\subsection{Depth of Observations}\label{sec:depth}
\begin{figure}
    \centering
    \includegraphics[width=.99\linewidth]{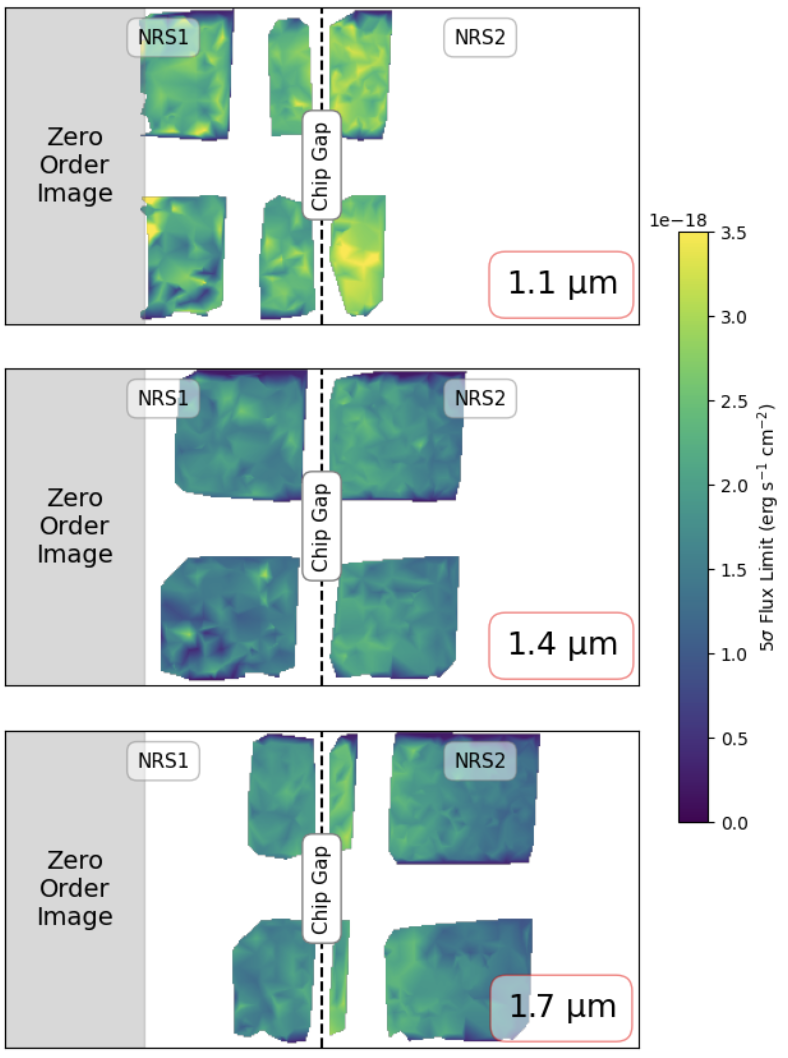}
    \caption{Wavelength slices of the depths of the MOCSS observations at $1.1\rm \mu m$ (top), $1.4\rm \mu m$ (middle), $1.7\rm \mu m$ (bottom) for one stack. The depth varies based on wavelength and spatial location on the detector, and the shallowest regions are typically near the chip gap. 
    }
    \label{fig:depthmap}
\end{figure}

The depth of observations varies both spectrally and spatially, in part due to the differing amount of open shutters in each row and column of the MSA. In order to investigate the typical depth of observations, we create a depth map by extracting the spectral traces for hundreds of equally spaced open microshutters from across all four quadrants of the MSA for stack A.  
From each trace, we compute the $5\sigma$ line flux limit (by calculating the $\rm \sigma_{rms}$ of the background and assuming a line width equivalent to that of the single line in MOCSSHUDF24) at different wavelengths for each trace, which corresponds to different spatial locations on the NIRSpec detector. We interpolate the depths for regions on the detector that are not directly covered by the selected extracted traces.
Example wavelength slices of the depth map can be see in Figure~\ref{fig:depthmap} for $1.1\rm \mu m$ (top), $1.4\rm \mu m$ (middle), $1.7\rm \mu m$ (bottom). 

The least sensitive locations on the NIRSpec detectors are the right side of NRS1 and the left side of NRS2. 
At its most sensitive, MOCSS observations can typically reach a $5\sigma$ flux depth of $\approx1.6 \times 10^{-18} \ \rm erg\ s^{-1} cm^{-2}$ in a single stack. For emission line identification (which is done with 2 stacks, see section~\ref{sec:identification}), the depth improves by a factor of $\sqrt{2}$. 
If the host source can be identified, the depth improves by a factor of $\sqrt{4}=2$ (reaching about $\sim8\times10^{-19}\ \rm erg\ s^{-1} cm^{-2}$). 
However, the emission line candidate selection with \textsc{sep} is done with only one stack at a time, and so this represents the objects that will be identified as legitimate emission lines in section~\ref{sec:identification}.

One important point to note here is that the true flux depth for MOCSS may be slightly shallower than reported: the spectroscopic flux calibration in the JWST pipeline is derived from observations of point sources and includes aperture correction for flux loss from the limited micro-shutter size in the dispersion direction. In the MOCSS observing mode, the slitless nature of these observations means that aperture-limiting effects are not susceptible to this limitation (given neighboring shutters are opened alongside one another in the dispersion direction). \citet{huberty2026mocssline} estimates this pipeline overestimate to be minor, and the flux loss is only about 3.3\%.

\subsection{UV Lines}\label{sec:uvlines}
The high-z UV lines of \ciii\ $\lambda1908.73$\AA, \civ\ $\lambda\lambda1548.20,1550.78$\AA\AA, \oiiib\ $\lambda1660.81$\AA, \heii\ $\lambda1640.42$\AA, \nv\ $\lambda\lambda1238.83,1242.80$\AA\AA, and \lya\ are of great interest for constraining the high-z Universe. At $z\gtrsim4.3$, these lines (starting with \ciii) begin to enter the MOCSS wavelength regime.

To explore the potential of detecting UV lines in light of the known depth of observations, we can verify from ancilliary spectroscopy whether we have captured all the emission lines that we should expect to see. 
An ideal survey for this is the JADES DR4 spectroscopic release\footnote{Also see \url{https://jades.herts.ac.uk/search/} for the public online database.} \citep{curtislake2025,scholtz2026}. 
JADES DR4 provides spectroscopic fluxes for the \ciii, \civ, \heii, and \oiiib\ lines, in particular measured from the NIRSpec medium resolution gratings similar to MOCSS. 
We also visually inspect the spectra and fits from the DR4 release for each of these potential lines, as there are occasionally overestimates in the line fluxes due to poor fits on weak or non-existent lines. 
We find that any sources with lines that would likely be strong enough to be seen in MOCSS have been masked, as they are too bright in the F814W (or in other words, at too low of a redshift). There are some UV-line emitting sources found in the DR4 catalog that would rise above the flux depth of MOCSS; unfortunately by chance they fall out of the MOCSS footprint. 

We repeat this exploration for all $z>7.25$ galaxies with JADES DR4 spectroscopic coverage that lie in the MOCSS footprint (16 of them are covered by all 4 MOCSS stacks). \lya\ falls in the MOCSS wavelength regime for $z>7.25$ sources (although it is worth mentioning that from a visual inspection of the JADES spectra, most do not exhibit strong \lya\ emission). In this manner, we do find a weak $z=8.48$, $\sim2.5\sigma$ \lya\ line as expected from the spectroscopy. We know its spectroscopic redshift (MOCSSHUDF23, this is not one of the \textsc{sep}-identified candidates, which had a detection threshold of about $\sqrt{5}\times3\sigma=6.7\sigma$). 
The spectrum for MOCSSHUDF23 can be seen in the final figure in the online figure set~\ref{figset:spectraexamples}.

Of course, JADES does not provide spectra of all our sources. Many strong UV line-emitters are likely missing from the targeted JADES spectroscopy that could appear in the MOCSS slitless observations. However, there is not enough evidence to classify any of the remaining unconstrained lines as such. 

A similar search for \lya\ can also be done with the First Reionization
Epoch Spectroscopically Complete Observations (FRESCO) survey \citep{oesch2023,meyer2024}. FRESCO is a NIRCam F444W survey that was able to spectroscopically constrain $6.9<z<9.0$ \oiii\ and \hb\ emitters. Unfortunately, only 4 FRESCO \oiii\ emitters are covered by all 4 stacks in MOCSS. Two of these sources are not at a high enough z to see \lya\ in the MOCSS regime, and in the other two sources, \lya\ lies right at the edge of the wavelength regime (meaning some of the spectral traces from each stack do not even cover it, and even if it does, the depth of observations is very low at the edge of the wavelength trace). These 4 sources also do not exhibit any noteworthy features at the expected locations of other strong UV lines. Therefore, no conclusions are made with the comparison to FRESCO.

\begin{figure*}
    \centering
    \includegraphics[width=.99\linewidth]{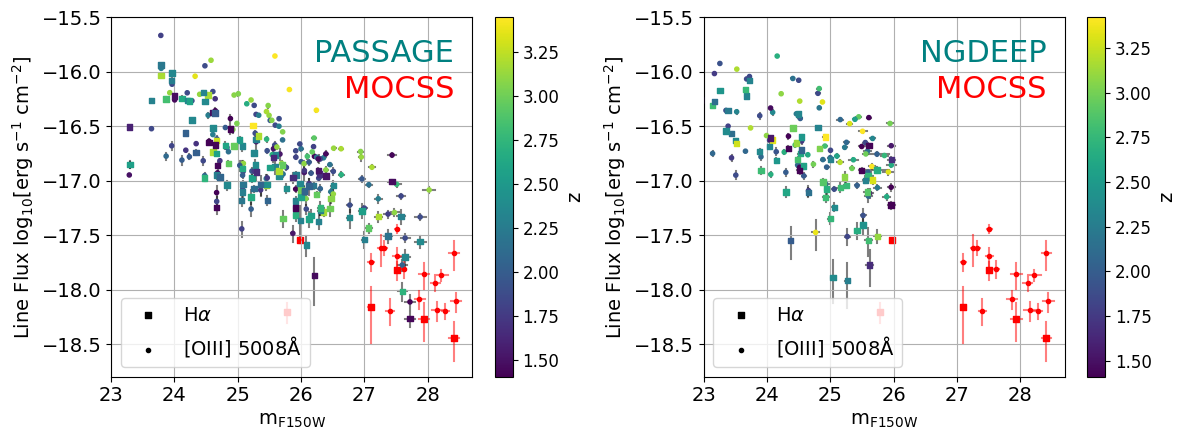}
    \caption{Comparison of \ha\ and \oiii\ 5008\AA\ line fluxes and F150W magntiudes for sources obtained in PASSAGE parallel field 028 \citep[left panel][]{huberty2026} and NGDEEP \citep[right panel][]{pirzkal2024,bagley2024} NIRISS WFSS (viridis colored points) and MOCSS (red points). The PASSAGE and NGDEEP points are colored by redshift, and are restricted to the same redshift regime as MOCSS. We find that MOCSS typically probes weaker line fluxes from fainter sources than both surveys.
    }
    \label{fig:surveycomp}
\end{figure*}

\subsection{Unconstrained Emission Lines}\label{sec:unknown}
Of 42 emission lines identified in Section~\ref{sec:identification}, 7 emission lines remain uncharacterized and unconstrained. 
These all appear to be single-line emitters, and have no partner emission line in the second orientation; therefore these lines remain without an identity, redshift, or host. 
While the lack of other emission lines could be used to further constrain these sources, there are a few issues with this: one is that these remaining lines tend to be weak. 
If the line is not very strong (for example, if the line is not strong enough to rule out \oiii\ 5008.24\AA\ line because the secondary \oiii\ 4960.30\AA\ line would fall below the level of the noise) it becomes nearly impossible to distinguish \oiii\ from \ha, \oii, \lya\ or any other potential lone emission line in the wavelength regime. Even if the line could be definitively identified as one of these lines, without the observation in the second orientation, the host becomes next to impossible to strongly constrain without at least one other line. If both the identity of a line and its observed wavelength was assumed, one could identify potential hosts with a similar JADES $\rm z_{phot}$ that would place the emission line at the expected wavelength. 
Nevertheless, the JADES $\rm z_{phot}$ cannot be considered a perfect estimate of the redshift even if they have often proven to be accurate (for an example with a large $\rm z_{phot}$/$\rm z_{spec}$ discrepancy, see MOCSSHUDF9 in Table~\ref{tab:mocsssources} in which the JADES $\rm z_{phot}$ strongly disagrees with both the MOCSS and JADES $\rm z_{spec}$). 
\citet{rauch2008} argues that many lone emission lines can be well explained by \lya, but without knowing the observed wavelength of the line, we cannot make any secure conclusions about potential redshifts of these lines.

\subsection{Comparison to Existing NIR WFSS}

We now investigate how the MOCSS survey compares to existing JWST NIR WFSS. JWST WFSS in the $1-2\rm \mu m$ regime is done with NIRISS. One comprehensive NIRISS WFSS survey is the Parallel Application of Slitless Spectroscopy to Analyze Galaxy Evolution survey \citep[PASSAGE,][]{malkan2025}. In particular, PASSAGE field Par028, provides F115W, F150W, and F200W WFSS at exposure times of 21.26, 7.08, and 5.37 hours respectively (each listed exposure time is divided equally into 2 orientations, i.e., one exposure of the F115W has an exposure time of $\sim10.6$ hours). This means that compared to a single MOCSS stack (in which each stack has an exposure time of $4\times2.33=9.34$ hours), the exposure time is similar to PASSAGE Par028 F115W in a single orientation (near 10 hours in each case). 

PASSAGE Par028 has well constrained line fluxes for sources in the same redshift range as MOCSS \citep[$1.4 \lesssim z\lesssim3.5$,][Nedkova et al. in prep]{huberty2026}. In the left panel of Figure~\ref{fig:surveycomp}, we compare the line fluxes for the \ha\ and \oiii\ 5008\AA\ emission lines with the F150W AB magnitude for both MOCSS (red points) and PASSAGE (viridis colored points). The F150 AB magnitudes for MOCSS are taken from JADES DR5, and the measurement of the F150W magnitude for both surveys is taken with an aperture diameter of 0\farcs5. We only plot the \textsc{sep} identified \ha\ and \oiii\ lines in MOCSS, even though fainter lines are identified in the final stacked spectra for each galaxy. Given that PASSAGE NIRISS observations have a lower resolution ($R\sim150$) than MOCSS, to differentiate the \oiii\ 5008\AA\ line flux from the reported blended \oiii 4960,5008\AA\AA\ line flux, we multiply the blended line flux by 2.98/(1+2.98) \citep{storeyzeippen2000}. We also only plot PASSAGE sources with robust, redshifts constrained by multiple strong emission lines \citep[``flag 1" sources in][]{huberty2026}.
We find that MOCSS survey typically probes just as deep (if not slightly deeper than Par028), obtaining weaker emission lines from fainter sources.

The advantage of MOCSS relative to NIRISS is that MOCSS gets WFSS spectroscopy for the full $1-2\rm \mu m$ wavelength regime without needing to use multiple filters (F115W, F150W, and F200W), which leads to a loss of wavelength coverage at about $1.3\rm \mu m$ and $1.7\rm \mu m$ due to filter gaps in NIRISS \citep[e.g.,][]{noirot2023,watson2025,pang2026,huberty2026}. In a similar overall exposure time (33.7 hours for PASSAGE Par028, 37.4 hours for MOCSS), the typical MOCSS line fluxes are as faint as the most extreme PASSAGE line fluxes.

We repeat this comparison in the right panel of Figure~\ref{fig:surveycomp}, but with NGDEEP NIRISS WFSS observations \citep{pirzkal2024,bagley2024}; a NIR WFSS survey that also covers the HUDF and partially overlaps with MOCSS. NGDEEP has exposure times of 52.2, 34.8, and 17.4 hours in the NIRISS F115W, F150W, and F200W filters respectively. 
It is essential to note that \citet{pirzkal2024} limits the extraction of NGDEEP spectra to sources brighter than $m_{AB,F160W}=26$, and therefore, the sources plotted in Figure~\ref{fig:surveycomp} are magnitude limited. 
However, even with this magnitude cut, the contrast between NGDEEP and MOCSS shows how the sources identified in MOCSS are very faint, without the need for multiples filters.
Together, these NIRISS WFSS comparisons demonstrate how MOCSS observations can provide an effective look at some of the faintest sources in the Universe.

\section{Conclusion}\label{sec:conclusion}
In this paper, we outline an experimental use of JWST/NIRSpec as a slitless spectrograph, known as Multi-Object Coronagraphy for Slitless Spectroscopy (MOCSS, JWST-GO-3290, PIs: Hayes and Scarlata, doi: \dataset[10.17909/tct8-jr92]{http://dx.doi.org/10.17909/tct8-jr92}). This technique involves opening essentially all the microshutters on the JWST/NIRSpec instrument to facilitate a slitless spectroscopy survey at $\rm R\sim1000$ in the $1-2\rm \mu m$ wavelength range. Simultaneously, shutters on top of sources identified in F814W imaging were closed, to prevent overlapping traces from bright sources from dominating the dispersed images.

The MOCSS survey of the HUDF/GOODS-S led to the identification of 42 emission lines, of which $\sim80\%$ have been robustly identified, alongside their redshifts and hosts in deep broad-band imaging. 
22 optical line emitters were spectroscopically identified at $1.395<z<4.537$. 18 of these spectroscopic redshifts are new  measurements and strongly agree with the $z_{\rm phot}$ from JADES DR5. The 4 sources with existing JADES spectroscopic redshifts also agree with those measured from MOCSS. 

We have also shown that the extended first order traces of NIRSpec/G140M/F100LP observation may contain substantial emission line coverage significantly passed the nominal upper wavelength of $1.89\rm \mu m$, as far as $\sim2.4-2.7\rm \mu m$.
We have also shown that this method has potential to identify even higher-z UV lines, but would benefit from even deeper observations. 

We also find that compared to existing NIR WFSS in the $1-2\rm \mu m$ regime, that MOCSS typically probes very weak emission lines ($\lesssim 1.6\times10^{-18}\rm\ erg\ s^{-1} cm^{-2}$) from faint sources ($\rm m_{F150W}\sim28$). 
These results highlight the discovery potential of this new observing strategy for WFSS and its promise for revealing some of the faintest and most elusive galaxies in the universe. Additionally, with the advent of MIRI WFSS in JWST Cycle 5, all 4 of the main instruments of JWST (NIRISS, NIRCam, NIRSpec, and MIRI) can now be used for WFSS surveys.

\section{Acknowledgments}
We thank the JWST help desk for providing substantial feedback and assistance with the data reduction.
We acknowledge support by NASA through grant JWST-GO-3290. All the JWST data used in this paper can be found in MAST: \dataset[10.17909/tct8-jr92]{http://dx.doi.org/10.17909/tct8-jr92}. M.S.H. thanks Andy Bunker for help with the JADES dataset.
M.J.H. is supported by the Swedish Research Council (Vetenskapr{\aa}det) and is Fellow of the Knut \& Alice Wallenberg Foundation.

\software{NumPy \citep{Numpy:2020}; SciPy \citep{Scipy:2020}; AstroPy \citep{Astropy:2013,Astropy:2018,Astropy:2022}; Matplotlib \citep{Matplotlib:2007}; SEP \citep{barbary2016,bertin1996}.}

\clearpage
\begin{appendix}

\section{Coronagraphy Preparation}\label{appenidx:maskpreparation}

The procedure to identify which NIRSpec shutters need to be closed to mask sources from F814W imaging began with the use of \textsc{source extractor} on F814W HUDF images \citep{whitaker2019}. A segmentation map of the field comprising of all sources with a SNR$\ge2$ and \textsc{minarea}$\ge5$ was created from \textsc{source extractor}. Bright compact galaxies were included as guide stars. This segmentation map was converted into an MSA catalog file for the JWST Astronomer's Proposal Tool (APT). The pointings were configured to ALLOPEN. 
Once an aperture position angle had been assigned, the central pointing was hand coded into APT and the target acquisition performed. The shutter configuration file was exported from the configuration editor, bypassing the MSA Planning Tool (MPT). The shutter configuration file is a 0-1 CSV file describing the shutter status for each microshutter. This file was inverted, so that all identified sources correspond to closed shutters, rather than open ones, and the newly updated file was re-imported to the APT. 
As a conventional dither is not achievable with this MSA configuration, this procedure needs to be repeated for all 8 configurations per visit.

\section{Exposure Information}\label{appendix:ditherinfo}
Table~\ref{tab:dithers} outlines information for each of the  dithers/exposures in the MOCSS survey. The NIRSpec observations are centered at R.A.=$53.16269^\circ$, Decl.=$-27.79192^\circ$. 
\begin{table*}[h]\label{tab:dithers}
\centering
\caption{Details of the MOCSS observations broken down by each exposure in the dither pattern. Each group of four exposures are aligned and stacked, resulting in stacks A, B, C, and D (see section~\ref{sec:observations}). The right ascension/declination offsets are reported with respect to the first exposure in each orientation (i.e., relative to exposure 1 for all orientation 1 objects; relative to exposure 9 for all orientation 2 objects). }\label{tab:dithers}
\begin{tabular}{cccccc}
\hline
Exposure             & Orientation          & Stack & Right Ascension Offset & Declination Offset  & Observation Date   \\
\multicolumn{1}{l}{} & \multicolumn{1}{l}{} &\multicolumn{1}{l}{} & ($^{\prime\prime}$)         & ($^{\prime\prime}$)               & \multicolumn{1}{l}{} \\ \hline
1                    & 1         & A           & 0                  & 0  & August 23, 2023                                      \\
2                    & 1          & A         & 0.26                             & 0.81   & August 23, 2023                                              \\
3                    & 1          & A          & 0.52                   & 1.61      & August 24, 2023                                  \\
4                    & 1         & A           & 0.77                   & 2.42             & August 24, 2023                          \\
5                    & 1         & B          & -15.97                  & 14.62      & August 24, 2023                                \\
6                    & 1          & B          & -15.71                         & 15.42         & August 24, 2023                                      \\
7                    & 1          & B          & -15.45                  & 16.23          & August 24, 2023                            \\
8                    & 1         & B           & -15.20                 & 17.03            & August 24, 2023                      \\
\hline
9                    & 2        & C            & 0                 & 0   & November 19, 2023                              \\
10                   & 2         & C           & 0.91                             & -0.24     & November 19, 2023                                      \\
11                   & 2          & C          & 1.81                  & -0.48          & November 19, 2023                           \\
12                   & 2         & C           & 2.72                   & -0.72             & November 19, 2023                     \\
13                   & 2          & D          & 16.79                  & 13.88          & November 19, 2023                   \\
14                   & 2           & D         & 17.70                         & 13.64     & November 19, 2023                                     \\
15                   & 2          & D          & 18.60                  & 13.40          & November 19, 2023                             \\
16                   & 2          & D          & 19.51                 & 13.15            & November 19, 2023                    \\ \hline
\end{tabular}
\end{table*}

\section{Dispersed Image File Modifications}\label{appendix:countrate}
The stacking of dispersion images require modifications of each extensions of the final stacked dispersed image. 
These modifications include:
\begin{itemize}
  \item The science extension (\textsc{sci}) of each stack is taken as the pixel by pixel weighted-mean of each exposure comprising that stack. 
  Each stack is aligned with the first exposure within that stack, as listed in Table~\ref{tab:dithers} (e.g., relative to expsoure 5 in stack B). This alignment also applies to every other extension of the fits files. 
  \item The data quality extension (\textsc{dq}) is modified so that if any of the four pixels that are stacked at each location in the substack have been flagged as a bad pixel, the pixel is given a $DQ=1$ (the signifier for DO\textunderscore NOT\textunderscore USE). Else, $DQ=0$. In the NRS1 detector stacks, all pixels that overlap with the zero order direct image of quadrants 1 and 2 of the NIRSpec/MSA are assigned $DQ=1$ (also see figure~\ref{fig:detectorimage} to visualize this overlap). 
  \item The \textsc{var\textunderscore poisson} and \textsc{var\textunderscore rnoise} variance extensions are replaced with the variance of a weighted average of the four individual variances in each stack. 
  \item The error extension (\textsc{err}) is overridden by the values of \textsc{var\textunderscore poisson} and \textsc{var\textunderscore rnoise} when the JWST pipeline is run. However, for initial visualization purposes of the stacked dispersion image error prior to spectral extraction, one can replace \textsc{err} with the pixel by pixel standard deviation of the stack.
\end{itemize}

\section{Details on Matching Emission Lines Between Orientations}\label{appendix:orientationmatchingcode}
When matching emission lines in the dispersed images from different orientations, the use of the JWST MSA metadata files and the JWST science calibration pipeline are used for the most accurate matching.
Given an emission line candidate, one can identify all MSA shutters that are open that could theoretically host the source of the emission line. For computational and temporal convenience, the original MSA metadata file hosting all shutter information can be culled to just the identified potential host shutters (also see Appendix~\ref{appendix:pipeline}). 

Then, the first step of the stage 2 JWST spectroscopic processing pipeline (i.e., \textsc{calwebb\textunderscore spec2}) should be ran: this step, \textsc{assign\textunderscore wcs}, attaches a world coordinate system transformation for each shutter between each pixel of the detector and the wavelength associated with that pixel given a spectral extraction from that shutter. Given that a shutter also corresponds to a location on the sky, this transformation effectively provides a transformation between sky location, wavelength, and pixel on the dispersed image. 

Therefore, this procedure is repeated for all emission lines identified by \textsc{sep}, where all potential host sky locations and their corresponding wavelengths are identified for all emission lines in both orientations. When emission line candidates from both orientations are attributed to the same sky location and wavelength, these sources are considered one in the same. This built-in transformation is not limited to the nominal NIRSpec/G140M/F100LP wavelength regime, but can extend well past the $1.89\rm \mu m$ red wavelength cutoff, which is useful in identifying emission lines in the extended first order trace. 

This transformation can be inverted to project all potential locations of an emission line identified in one orientation into the dispersed image of the other orientation. This is particularly useful in section~\ref{sec:3stacks}, in order to visually confirm the location of emission line in the second orientation when it was not identified by \textsc{sep} in both stacks of this second orientation.

\section{JWST MSA Metadata File and Pipeline Modifications}\label{appendix:pipeline}
The JWST MSA metadata file that is associated with every NIRSpec observation contains crucial information regarding the state of every microshutter and source configuration. 
In typical MOS data reductions, the JWST pipeline will extract the spectra for each open microshutter identified in the MSA metadata file. This is not feasible for MOCSS observations, not least because of computational reasons: this would require the extraction of 100,000s of spectra (one for every open shutter) in just one single stack. A majority of these spectra would not contain any features of interest, and given the length of the G140M/F100LP traces, many overlapping spectral extraction would include the same potential emission line hundreds of times. This would not majorly assist in the identification of which host an emission line belongs to.

Therefore, following the identification of sources in section~\ref{sec:sourceidentification}, modifications to the original MSA metadata files are essential for MOCSS spectral extractions. The following modifications are made: 
\begin{itemize}
  \item The MSA metadate file is culled to remove all shutters (identified by \textsc{shutter\textunderscore row} and \textsc{shutter\textunderscore column} coordinates) that are not of interest. 
  \item Each dither has a specific \textsc{msa\textunderscore metadata\textunderscore id} identifier. The MSA metadalta file is further culled to exclude all shutters that do not have the appropriate \textsc{msa\textunderscore metadata\textunderscore id} for the dither (or stack) of interest, otherwise it will extract the spectrum from the wrong part of the detector. 
  \item If extracting more than one shutter at a time, the slitlet identification numbers (\textsc{slitlet\textunderscore id}) must be renamed unique identifiers for a very important reason: a `slitlet' in typical MOS observation refers to a small group of shutters that are assigned to one source in the field, all of which contain the source through the dither/nod pattern. 
  The shutters of one slitlet lie next to each other in the cross-dispersion direction. By opening multiple neighboring shutters in the cross dispersion direction, every grouping of shutters in the cross-dispersion direction are assigned the same \textsc{slitlet\textunderscore id}. This results in the extraction of spectral traces (in the cross-dispersion direction) that are as thick as the number of shutters (in fact, if there are no closed shutters in the same column of the MSA as the shutter being extracted, the entire dispersion image can be cutout as a single trace without this renaming or culling of shutters, which is nonsensical). 
  This renaming is also useful for book keeping, by assigning \textsc{slitlet\textunderscore id} based on the individual shutter location in the field of view, it is easier to identify the shutter in one orientation/dither that corresponds to the same location in the field of view as the other orientation/dithers. 
\end{itemize}

These metadata changes apply to spectral extractions for both the \citet{jwstpipeline2023} JWST science calibration pipeline and the \citet{msaexp2022} manual NIRSpec extraction pipeline. With these metadata modifications in place, the spectral extraction of the stage 2 JWST pipeline can be run. 
The stage 2 pipeline applies instrumental corrections and calibrations to the count-rate products that result in fully calibrated exposures. The pipeline steps of \textsc{master\textunderscore background, pathloss,} and \textsc{barshadow} are skipped, due to them not being required in a slitless spectroscopic campaign or the dither pattern making them unneeded. 
Running this part of the pipeline results in calibrated science products, alongside 2D and 1D spectral products. 

The stage 3 JWST pipeline combines data from multiple exposures into a single countrate product. As we account for multiple exposures in a different manner than for typical NIRSpec/MOS data, we pass on this portion of the pipeline.

\end{appendix}

\bibliography{refr}{}
\bibliographystyle{aasjournalv7}

\end{document}